# An investigation on the effect of the near-fault earthquakes on the seismic behavior of RC Moment Resisting Frames (MRFs) designed based on Iranian seismic code (standard no. 2800)

Mojtaba Harati[1]

Ahmad Hojjati[2]

Ataalh Modaraei[3]

**Abstract**

Past severe earthquakes, such as the Bam earthquake of 2003 and Tabas earthquake of 1978, have demonstrated that many cities in Iran are prone to be struck by near-fault earthquakes. Such earthquakes are impulsive in nature, and therefore, they are more destructive than the ordinary ground shakings. In the fourth edition of Iranian seismic code (Standard No. 2800), some changes, including a modification factor for the elastic acceleration response spectrum (EARS) have been recently recommended to reflect the effects of such probable near-fault earthquakes for the designing procedure. In this study, a number of 2D RC moment resisting frames (MRFs), from four- to a twelve-story building, is designed linearly based on Iranian National Building Code (INBC) and Standard No. 2800 as well. Subsequently, their nonlinear models are reproduced for conducting a nonlinear dynamic time history (NDTH) analysis. For this purpose, twenty impulsive ground motions are selected and scaled to be compatible with the design basis earthquake (DBE) spectrum of the abovementioned codes. It is concluded that the seismic performance of the analyzed structures is not satisfactory at all; no buildings are successful to satisfy the life safety (LS) performance level posed by guidelines such as ASCE41-06 or ASCE41-13. Moreover, it is worth mentioning that even collapse prevention (CP) limit states are not also met in some cases. Therefore, the recently added modifications in Standard No. 2800 may be inadequate to incorporate the near-fault earthquakes' effects. In the end, some recommendations are addressed for the design of structures built in the regions vulnerable to be attacked by near-fault earthquakes.

---

1 Lecturer, University of Science and Culture, Moj.harati@gmail.com

2 Msc student of structural engineering, Azad University of Anzali

3 Assistant Professor, Department of Civil Engineering, University of Guilan

## Introduction

Earthquake is a natural phenomenon which kills many people annually all over the world. It happens when a fault initiates to get ruptured. In this case, a huge amount of elastic energy, potentially available in the rocks, is released from the source of earthquakes when a point in a specific slip starts faulting. As the fault continues to rupture, and this resulted relived energy goes through the layers of the earth, generated elastic waves travel towards the areas where people have selected to live. According to the relative distance of these areas to the source of earthquakes, two kinds of earthquakes are usually generated. When the epicenter of the earthquakes is within 15 (Ambraseys and Douglas 2003) to 20 (Bray and Rodriguez-Marek 2004) kilometer from a struck area, they are called near-fault earthquakes. On the other hands, when the areas are far away from the epicenter of the ground shakings, they are called far-fault earthquakes. In general, near-fault earthquakes are more destructive than the far-fault ones, because they are closer to the sites they attack, and actually, their energy are normally less dissipated (Naeim 1995). As a result, these kinds of earthquakes are always associated with higher peak ground velocities (PGVs) and shorter strong motion duration (Bray and Rodriguez-Marek 2004). The differences of these two earthquakes were noticed by several researches in the past (Naeim 1995, Chopra and Chintanapakdee 2001, Hall et al. 1995).

A lot of cities have been constructed and developed on or near the active faults all over the world. In Iran, many active faults go through the populated cities like Tehran, Mashhad and Bam to name a few. The past severe near-fault earthquakes, such as Kobe earthquake of 1995 and Tabas earthquake of 1978 (Yaghmaei-Sabegh and Tsangb 2011), have shown that many cities in the world have such a critical situation. Therefore, many seismic codes all over the world have been revised to include the effects of such near-fault earthquakes on the seismic design of their infrastructures. In such way, Iranian seismic code (4$^{th}$ edition of Standard No. 2800) has been recently revised to incorporate the effect of such near-fault earthquakes.

RC structures are being utilized as structural systems in construction industry worldwide. They are also abundant in Iran because of their compatibility with the environmental capacity of this country. In particular, RC moment resisting frame (MRF) is one of the most popular and common structural systems in Iran. In this investigation, the seismic performance of RC MRFs, designed based on the fourth edition of Iranian seismic code (Standard No. 2800), is studied thoroughly. In this case, three 2-dimentional RC MRFs, four to twelve-story buildings, are designed with the recent edition of Iranian seismic code (4th edition of Standard No. 2800). Subsequently, their nonlinear models are generated in Seismostruct v7.2 software (Seismosoft 2013) for performing a set of NDTH analysis. Moreover, it is also worth mentioning that a set of near-fault earthquakes are selected as input ground motions for the NDTH analysis. Therefore, the results of the analysis are presented within the performance based design (PBD) method framework. Therefore, the adequacy of recently added provisions in Iranian seismic code (Standard No. 2800) to include the effects of the near-fault earthquakes are assessed via NDTH analysis using natural impulsive ground motions.

## Near-fault earthquakes

Areas and regions, where experience near-fault earthquakes, are divided into three zones. zones where their ground motions is influenced by 'Forward Directivity', 'Backward Directivity' and 'Neutral Directivity' as well. Just one of these zones experience impulsive earthquakes that cause structures to get heavily damaged. This area is where the forward directivity happens. Forward directivity is a natural phenomenon in which the direction of the earthquake's wave propagation is coincidence with the direction of the slip faulting (Somerville et al. 1997). Moreover, in this phenomenon, the velocity of earthquake's wave propagation and the faulting is nearly the same (Bray and Rodriguez-Marek 2004). In this case, the sites toward which slips, strike-slips or dip-slips, rupture is called forward directivity zone. To be specific, the region within the angles of faulting direction and the path through which earthquakes' wave goes ahead is where the forward directivity effects are expected to occur (Bray and Rodriguez-Marek 2004). This angle is small in nature and is shown in Figure 1. On the other hand, in the opposite direction to which the forward directivity happens, earthquakes are not generally impulsive at all. Additionally, regions that are located away from the aforementioned locations are expected to experience a neutral directivity effect in their ground motions which is nearly the same as the ground motions of far-fault earthquakes.

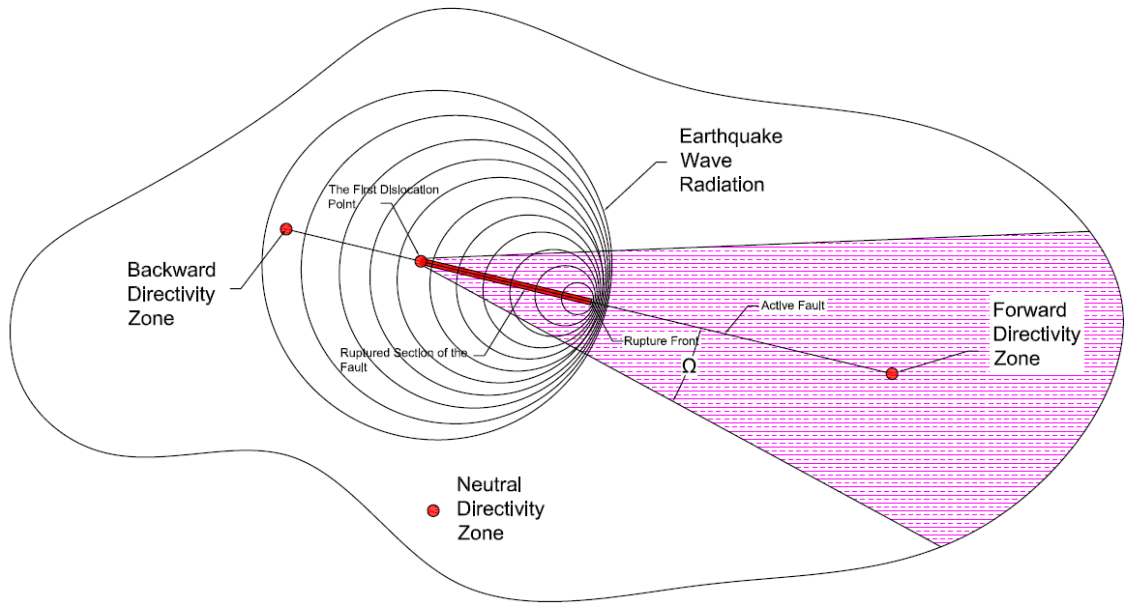

**Figure 1,** forward, backward and neutral directivity zones in a region near to an active fault.

Ground motions in this zone are also called forward directivity (FD) ground motions. Big pulses with high amplitude and intensity with a period of 1.5 to 3 sec are always apparently seen in the velocity and displacement time histories of the fault normal-direction of the FD ground motions (FDGMs). However, strong pulses may be existed in the fault parallel-direction as well (Bray and Rodriguez-Marek 2004). Occasionally but not always, these pulse-like shapes are also recognizable in the acceleration time history of the FDGMs. So, there are two distinctive characteristics in the ground motions of the near-fault earthquakes influenced by forward directivity. First, these ground motions are always followed by immense peak ground velocities (PGVs) in one of those a few pulse-like shapes in their velocity time histories. Second, the earthquakes influenced by forward directivity have a shorter strong motion duration compared to the ordinary quakes; because the pulses usually existed at the onset of these quakes, induce a huge amount of shakings' energy to the human-made structures. Therefore, these huge amounts of energy, which is cumulative in nature and released within a short period of time, are correspondingly followed by an extraordinary PGV both of which create severe hurtful effects on the infrastructures.

**Ground Motions Selection**

   Time histories nonlinear analysis is the method used for structural analysis of the case study buildings. For this purpose, near-fault earthquakes with return period of 475 years have been selected to be used. To select these earthquakes, scaling methods are usually used to match earthquakes acceleration response spectrum to predefined spectra. If these earthquakes are supposed to be selected with this approach, design based earthquake (MCE) response spectra or design basis earthquake (DBE) spectra may be used as the predefined spectra. In this method, intensity-based scaling of ground motion records using appropriate scale factors is required so that the mean value of the 5%-damped response spectra for the set of scaled records is not less than the predefined response spectrum (such as MCE or DBE) over a suitable range of period. There are different structures with different fundamental periods in this study, therefore, a wide range of fundamental periods are encountered. In this case, a set of records whose response spectrum covers the whole range of these fundamental periods is desirable. Based on the fact that the procedure to collect the acceptable ground motion records is a sensitive and professional task of seismologists, near-fault ground motion records of SAC steel project have been selected to be applied for the analysis. Information related to the selected near-fault earthquakes are condensed in Table 1. It is worthwhile to mention that DBE spectrum of fourth edition of Standard No. 2800 is employed to be the target predefined spectrum.

Table 1, near-fault earthquakes of SAC steel project

| SAC Name (Earthquake ID) | Record | Earthquake Magnitude | Distance (km) |
|---|---|---|---|
| NF01 | Tabas, 1978 | 7.4 | 1.2 |
| NF02 | Tabas, 1978 | 7.4 | 1.2 |
| NF03 | Loma Prieta, 1989, Los Gatos | 7 | 3.5 |
| NF04 | Loma Prieta, 1989, Los Gatos | 7 | 3.5 |
| NF05 | Loma Prieta, 1989, Lex. Dam | 7 | 6.3 |
| NF06 | Loma Prieta, 1989, Lex. Dam | 7 | 6.3 |
| NF07 | C. Mendocino, 1992, Petrolia | 7.1 | 8.5 |
| NF08 | C. Mendocino, 1992, Petrolia | 7.1 | 8.5 |
| NF09 | Erzincan, 1992 | 6.7 | 2 |
| NF10 | Erzincan, 1992 | 6.7 | 2 |
| NF11 | Landers, 1992 | 7.3 | 1.1 |
| NF12 | Landers, 1992 | 7.3 | 1.1 |
| NF13 | Northridge, 1994, Rinaldi | 6.7 | 7.5 |
| NF14 | Northridge, 1994, Rinaldi | 6.7 | 7.5 |
| NF15 | Northridge, 1994, Olive View | 6.7 | 6.4 |
| NF16 | Northridge, 1994, Olive View | 6.7 | 6.4 |
| NF17 | Kobe, 1995 | 6.9 | 3.4 |
| NF18 | Kobe, 1995 | 6.9 | 3.4 |
| NF19 | Kobe, 1995, Takatori | 6.9 | 4.3 |
| NF20 | Kobe, 1995, Takatori | 6.9 | 4.3 |

## Case Study Buildings

**Designing method**

There are two approaches to incorporate the effects of near-fault earthquakes in the design of infrastructures. In the first approach, near-fault effects are indirectly taken into account by modifications to the elastic acceleration response spectrum (EARS) at 5% damping (Somerville et al. 1997, Somerville 2003, Spudich and Chiou 2008). In this approach, the shape of EARS is amplified by a number which is mainly a function of structures' period. In this way, the function of EARS is intensified from low to long period regions. As can be seen in Figure 2, Iranian seismic code (Standard No. 2800) has been recently revised to incorporate the effect of such near-fault earthquakes by multiplying a function to the EARS for mid to long periods range. It is also worth mentioning that this function is named N factor which is mainly a function of structures' period. For the second approach, the effects of near-fault earthquakes are directly taken into account by performing a nonlinear dynamic time history (NDTH) analysis using near-fault earthquakes as input ground motions (Mavroeidis and Papageorgiou 2003, Bray and Rodriguez-Marek 2004, Baker 2007). While some researchers say that the first way is adequate to incorporate the effects of near-fault earthquakes (Malhotra 1999, Chopra and Chintanapakdee 2001), there are many other ones who endorse the next way (Anderson and Bertero 1987, Hall et al. 1995, Makris 1997, Alavi and Krawinkler 2000, Sasani and Bertero 2000, Mylonakis and Reinhorn 2001, and Zhang and Iwan 2002, Sehhati et al. 2011). The latter group believes that the seismic response of the structures under impulsive earthquakes is better captured with the NDTH analysis.

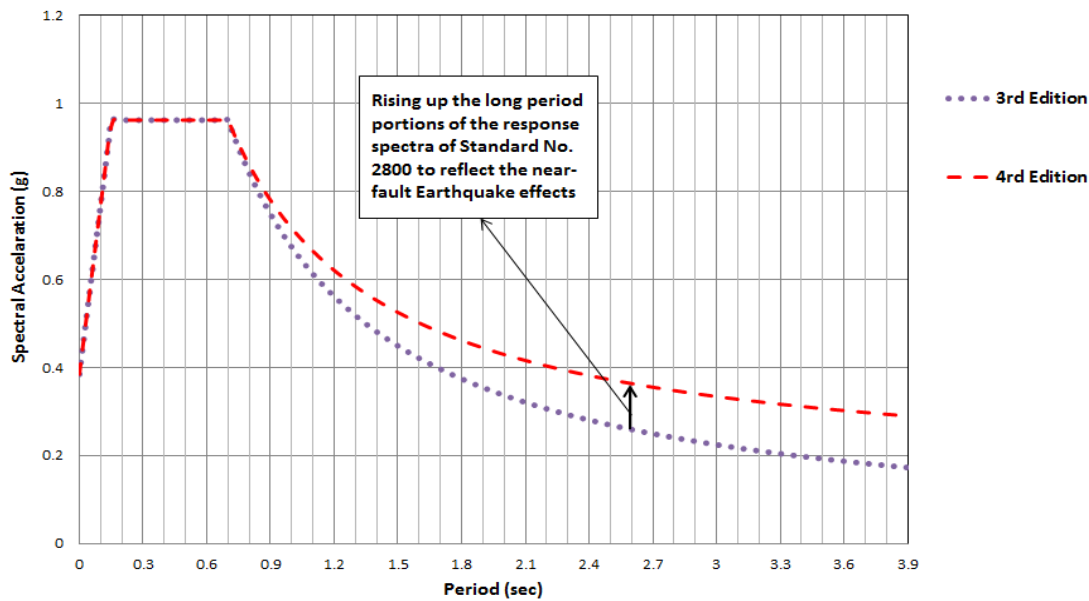

**Figure 2,** Rising up the long period portions of the response spectra of Standard No. 2800 to reflect the near-fault Earthquake effects

### Designing process

MRFs of case study buildings are designed linearly based on provisions of Iranian National Building Code (INBC) and fourth edition of Standard No. 2800. General geometry properties of the buildings, including span lengths, story heights and the numbers of stories have been carefully chosen in such a way that they would be a good representative of the values used in common construction practice of Iran. Furthermore, the concrete compressive strength and the reinforcement yield strength, used in the design procedure, are about 21 and 400 Mpa respectively. These material properties have been selected from which materials characteristics that was employed to verify the tested RC MRF structures (Varum, H 2003) model in Seismostruct software (Seismosoft 2013). To carry out a parametric study, low to mid-rise buildings with four, eight, and twelve stories have been designed through equivalent static lateral force procedure. The details of these structures are presented in tables 2, 3 and 4 respectively.

**Table 2,** Member sizes and structural details for four story buildings

| Story | Beams-b x h ($\rho^T$ (%) _ $\rho^B$ (%)) | Columns-b x h ($\rho$ (%)) |
|---|---|---|
| 1 | 40x40 (0.67-0.30) | 45x45 (1.07) |
| 2 | 40x40 (0.77-0.32) | 40x40 (1) |
| 3 | 35x35 (0.91-0.29) | 40x40 (1) |
| 4 | 30x30 (1.1-0.32) | 35x35 (1.15) |

**Table 3,** Member sizes and structural details for eight story buildings

| Story | Beams-b x h ($\rho^T$ (%) _ $\rho^B$ (%)) | Columns-b x h ($\rho$ (%)) |
|---|---|---|
| 1 | 55x55 (0.32-0.17) | 60x60 (1.3) |
| 2 | 55x55 (0.33-0.17) | 60x60 (1.2) |
| 3 | 50x50 (0.37-0.30) | 55x55 (1.1) |
| 4 | 50x50 (0.47-0.29) | 55x55 (1) |
| 5 | 45x45 (0.46-0.30) | 50x50 (1) |
| 6 | 40x40 (0.59-0.30) | 45x45 (1) |
| 7 | 35x35 (0.69-0.29) | 40x40 (1) |
| 8 | 30x30 (0.84-0.28) | 35x35 (1.15) |

**Table 4,** Member sizes and structural details for twelve story buildings

| Story | Beams-b x h ($\rho^T$ (%) _ $\rho^B$ (%)) | Columns-b x h ($\rho$ (%)) |
|---|---|---|
| 1 | 60x60 (0.36-0.23) | 65x65 (1) |
| 2 | 60x60 (0.36-0.23) | 65x65 (1) |
| 3 | 60x60 (0.36-0.23) | 65x65 (1) |
| 4 | 55x55 (0.43-0.32) | 60x60 (1) |
| 5 | 55x55 (0.43-0.32) | 60x60 (1) |
| 6 | 50x50 (0.51-0.33) | 55x55 (1) |
| 7 | 50x50 (0.52-0.32) | 55x55 (1) |
| 8 | 45x45 (0.57-0.32) | 50x50 (1) |
| 9 | 45x45 (0.55-0.32) | 50x50 (1) |
| 10 | 40x40 (0.65-0.30) | 45x45 (1) |
| 11 | 35x35 (0.73-0.29) | 40x40 (1) |
| 12 | 30x30 (0.83-0.28) | 35x35 (1) |

**Nonlinear modeling procedure**

Nonlinear models required for the NDTH analysis have all been recreated using Seismostruct v7.2 software (Seismosoft 2013). Hence, all the members of 2D structures are modeled through the fiber section method technique. Both the elements types and the material inputs of the fibers are compatible with the models' characteristics of which tested structures that are verified analytically in the software (Panagiotou et al. 2006). In terms of elements types, inelastic frame element with concentrated plasticity is employed for beams and columns modeling. In terms of material inputs of the fibers, Mander (Mander et al. 1998) and Menegotto (Menegotto and Pinto 1973) nonlinear input characteristics are selected for materials of the concrete and steel fibers correspondingly. It is also noteworthy to say that in nonlinear modeling procedure, floors are assumed to be rigid diaphragms, but the slabs participation are taken into accounts considering effective width beam, recommended by American Concrete Institute (ACI) (2011).

The generated models are subsequently used for performing the NDTH analysis. The direct integration of the equations of motion is accomplished using the numerically dissipative integration algorithm (Hilber et al. 1977) with automatic time-step adjustment for optimum accuracy and efficiency. Seismic masses are taken as the dead loads plus 30% of the live loads, based on the provision of ASCE07 (2013) for conducting NDTH analysis in the residential buildings. Furthermore, a 5% tangent stiffness proportional damping (Priestley and Grant 2005), which is appropriate for RC buildings, is employed throughout this investigation.

**Seismic Performance Criteria**

To evaluate seismic performance of a building, suitable seismic indicators are of essence. In this investigation, seismic performance of the buildings in near-fault earthquakes is aimed to be assessed. In this case, appropriate seismic indicator, such as the ones corresponding to life safety or near-collapse limit state may be required. Members' Plastic hinge rotation and maximum nonlinear inter-story drift ratio (IDR) have been designated as a seismic performance indicator for both local, for the structural members, and global scale respectively. In the case of plastic hinge rotation, life safety (LS) and collapse prevention (CP) performance level based on ASCE41-13 standard seems to be suitable. Two and four percent maximum inter-story drift ratio are also appropriate for the global seismic performance limit state, the first percent is based on ASCE07 design earthquake limit state, and the latter is based on ASCE41-13 standard for the CP performance level.

**Results**

**Nonlinear IDRs and rotational demands**

In this section, the results of nonlinear inter-story drift ratios (IDRs) and elements' rotational demands of the case study buildings are presented. For this purpose, nonlinear dynamic time history (NDTH) analysis is employed using twenty selected impulsive ground motions. Both components of the selected near-fault earthquakes in the horizontal surface are selected to be used for conducting NDTH analysis. This is mainly due to the nature of such quakes; just one of these components may show a forward directivity effect (Bray and Rodriguez-Marek 2004). Unlike far-fault ground motions, it is worth mentioning that near-fault ground motions have an intensive vertical ground motion component (Bray and Rodriguez-

Marek 2004). Therefore, vertical components of the selected ground motions are also employed for this investigation.

In the case of IDRs, the distributions of the IDRs are demonstrated alongside the height of the structures in Figure 3 and 4. In these figures, the horizontal and vertical axis represents the nonlinear IDRs in percent and the number of stories respectively. Also, Blue lines represent the results of each individual applied near-fault ground motions. In this case, each blue line is an output of a NDTH analysis independently. In general, in the lower portions of the structures, there is an increase in the IDRs, reaching to a peak value in one of those stories located in the bottom of the structures. In the RC MRFs, these mentioned concentrated IDRs in the bottom stories occur mainly due to the P-$\Delta$ effects of such frames (Moghaddam and Harati 2015). As can be seen from the blue lines, the structures' responses vary significantly through all applied ground motions. This shows that the behavior of such frames is highly sensitive to the nature of the selected near-fault ground motions. Moreover, red line also shows the average response of the structures as well. As can be seen, the data available from the blue lines are highly scattering, therefore, the average response does not seem to be appropriate solely. As a result, an average plus one standard deviation response is displayed in these Figures.

In four story buildings, which are good representative of common low-rise RC buildings in construction industry of Iran, the seismic behavior is not appropriate at all. As can be seen in Figure 3, nearly in all of the applied near-fault earthquakes, it is unable to maintain a life safety (LS) performance level; many blue line curves cut the LS limit state of the IDRs of RC MRFs posed by ASCE41-06. It is worthwhile to repeat that all selected earthquakes are within the DBE hazard level, and therefore, designed buildings are anticipated to meet the LS limit state in this case. As can be seen, red and green lines also cut the LS limit state. The interesting fact is that in several cases, this building is not also able to satisfy the collapse prevention (CP) performance level too. In this way, the average plus one standard deviation response is also about to touch the CP limit state. Therefore, this building is in danger of getting collapsed in some cases.

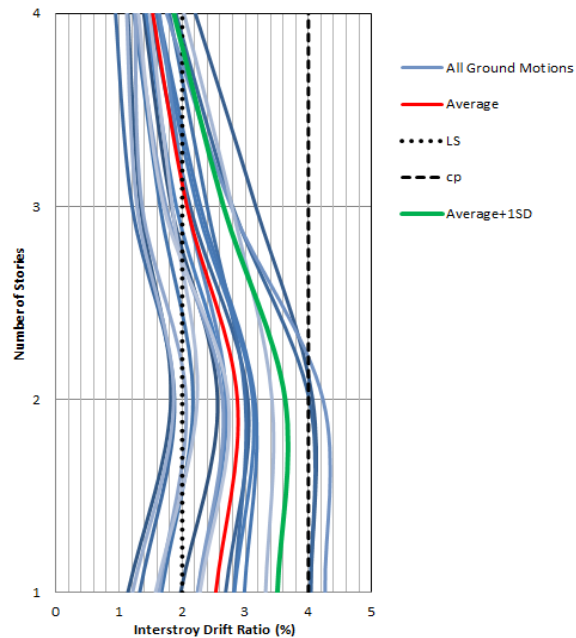

**Figure 3,** Inter-story drift ratio's profile for the four story building

In eight and twelve story buildings, which are good representative of common low to mid-rise RC buildings in construction industry of Iran, the seismic behavior is not suitable too. In many of the NDTH analysis via near-fault earthquakes, they are not capable of keeping a life safety (LS) performance level; many blue line curves intersect the LS limit state of the IDRs of RC MRFs recommended by ASCE41-06. It is worthy to mention that all selected impulsive quakes are within the DBE hazard level, and therefore, designed structures are expected to readily meet the LS limit state in this case. As can be seen in figure 4, red and green lines also surpass the LS limit state. The remarkable fact is that in several cases, these buildings are not also capable of satisfying the collapse prevention (CP) performance level as well. In this regard, the average plus one standard deviation response is also about to exceed the CP limit state. Therefore, these buildings are in danger of getting collapsed in some cases.

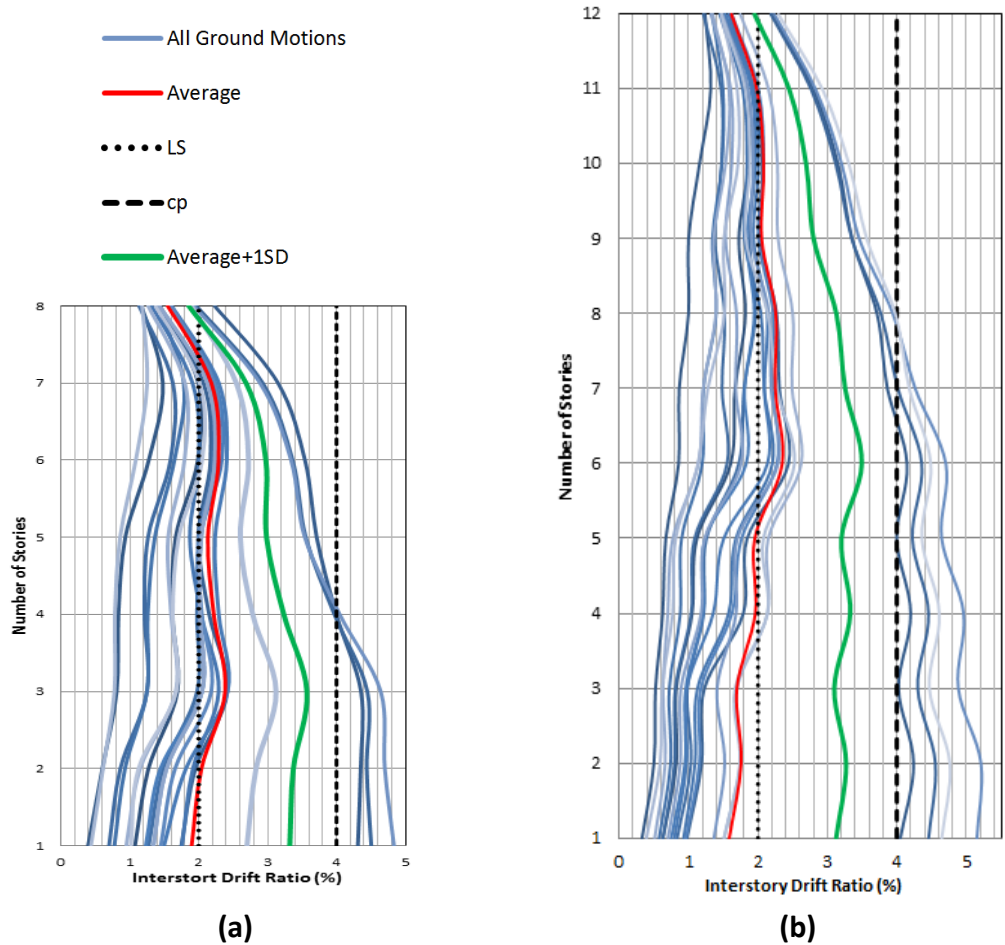

**Figure 4,** Inter-story drift ratio's profile for (a) eight story building and (b) twelve story building

Beside the results of IDRs that have been presented in previous section, the results of rotational demands of the structural members, including the beams and the columns rotation, are also presented here within the Performance Based Design (PBD) framework. Two different performance levels, namely the LS and the CP rotational based limit states, are considered for this stage of the investigation. Information and the results of these PBD analyses for all near-fault ground motions are condensed in table 2.

**Table 2,** seismic performance status of the case study buildings in the PBD framework

| SAC Name (Earthquake ID) | Four Story Building | | Eight Story Building | | Twelve Story Building | |
|---|---|---|---|---|---|---|
| | **LS** | **CP** | **LS** | **CP** | **LS** | **CP** |
| NF01 | Reached | - | - | - | - | - |
| NF02 | Reached | - | - | - | - | - |
| NF03 | Reached | Reached | Reached | Reached | Reached | Reached |
| NF04 | Reached | Reached | Reached | - | Reached | - |
| NF05 | Reached | Reached | Reached | Reached | Reached | Reached |
| NF06 | Reached | - | Reached | - | Reached | - |
| NF07 | Reached | Reached | - | - | - | - |
| NF08 | Reached | - | - | - | - | - |
| NF09 | Reached | - | Reached | - | Reached | - |
| NF10 | Reached | Reached | Reached | - | Reached | - |
| NF11 | Reached | - | Reached | - | - | - |
| NF12 | Reached | - | Reached | - | - | - |
| NF13 | Reached | Reached | Reached | Reached | Reached | Reached |
| NF14 | Reached | - | Reached | - | Reached | - |
| NF15 | Reached | - | - | - | - | - |
| NF16 | Reached | - | - | - | - | - |
| NF17 | Reached | Reached | Reached | - | Reached | - |
| NF18 | Reached | - | Reached | Reached | Reached | - |
| NF19 | Reached | - | Reached | - | Reached | Reached |
| NF20 | Reached | - | Reached | - | Reached | - |

As it can be seen from table 2, for four story building, for all applied near-fault ground motions, LS performance level is reached. This means that in this category, probability of not satisfying the LS limit state is 100% and therefore, low rise MRFs are in most critical status compared with the ones studied in this investigation. The probability of being LS for the eight and twelve stories buildings is 70% and 60% correspondingly. Therefore, the seismic performance of the eight and twelve stories buildings is a bit better. Additionally, in several cases for all categories investigated in this study, the CP performance level is not met as well.

**Discussions**

Although no buildings designed based on Standard No. 2800 demonstrate a satisfactory seismic behavior, taller buildings are more successful to curb the IDRs in the near-fault earthquakes. In figure 5, maximum IDRs for the average, average plus one standard deviation and the maximum response of all investigated buildings are depicted. As can be seen, a decreasing trend is obviously seen in the average and average plus one standard deviation response of the structures respectively. Therefore, taller the structures are, much better the seismic performance will be. This demonstrates that low rise buildings are highly susceptible to near-fault earthquakes while they are designed based on Standard No. 2800.

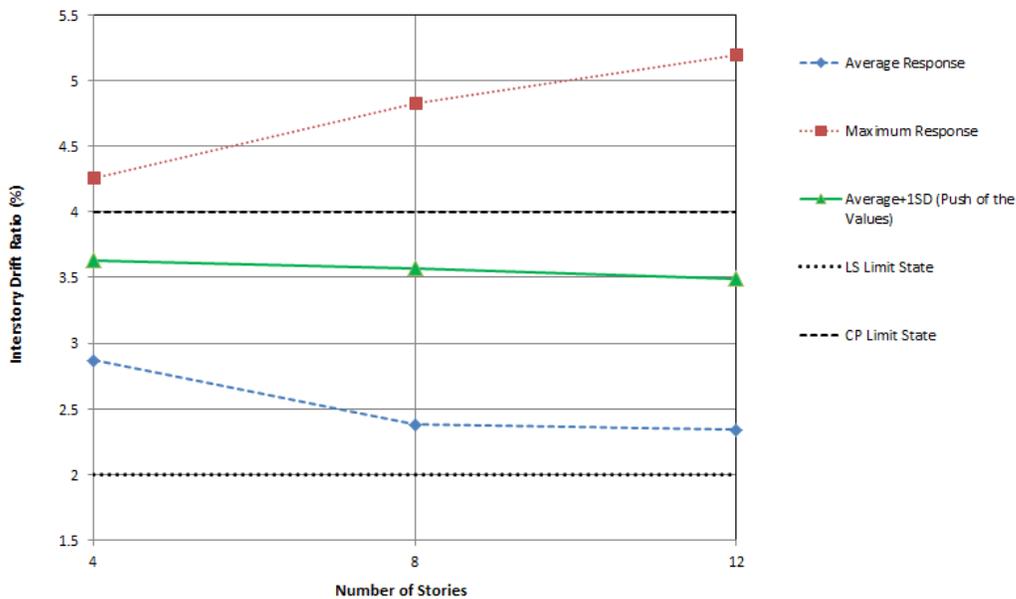

**Figure 5,** variation of maximum IDRs with number of stories

To scrutinize the behavior of the buildings designed based on the latest version of Standard No. 2800, a spectral based approach is used here to assess the seismic performance of such MRFs qualitatively. In figure 6, the average EARSes of far and near-fault ground motions are depicted beside the standard design spectra of the third and fourth edition of Standard No. 2800. In this case, far-fault ground motions and near-fault ones are selected from FEMA P695 guideline (2009) and SAC project respectively. Subsequently, they are scaled to the design spectrum of fourth edition of Standard No. 2800 for the period range of a four story RC building. For scaling procedure, the PGA of the design spectra of Standard No. 2800 is scaled to be unit (g m/s$^2$).

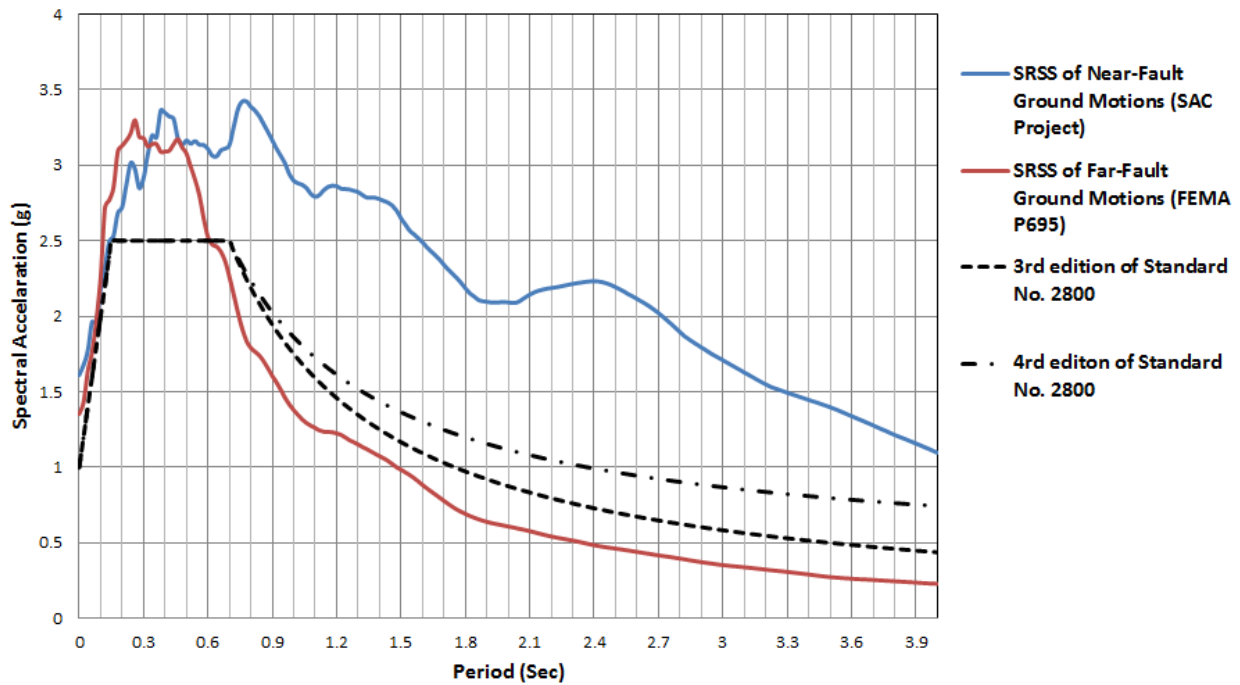

**Figure 6,** SRSS of Near-Fault ground motions, SRSS of Far-Fault ground motions, standard design spectra of the 3$^{rd}$ and 4$^{rd}$ edition of Standard No. 2800

In the case of near-fault earthquakes, the average spectrum is far above the standard design spectrum of the fourth edition of Standard No. 2800. Therefore, in the mid to long period regions, the recently modified section of the design spectrum seems to be ineffective. In general, if the buildings near to an active fault are aimed to be designed with linear based methods, the standard design spectrum recommended by standard No. 2800 does not seem to be proper. As a result, for designing the buildings located near to an active fault, it is recommended to generate a site response spectrum. Site response spectrum is usually obtained from a seismic hazard analysis (SHA) of the faults existed around the site buildings are determined to be built. However, the best way for designing buildings which are decided to be erected near to an active fault is using PBD framework. In this case, a bunch of near-fault ground motions compatible with the soils condition of the site should be selected for the NDTH procedure by highly qualified seismologists.

In the case of far-fault earthquakes, the average spectrum is far below the standard design spectrum of the fourth edition of Standard No. 2800. So, in the mid to long period regions, the recently revised section of the design spectrum looks to be overestimating. In general, if the buildings far to an active fault are designed with linear based methods, the designed structures sound to be overdesigned.

## Conclusions

Many places in Iran, including large populated cities such as Tehran, Mashhad and many other ones, are prone to be attacked by near-fault earthquakes. In this case, Manjil-Rudbar earthquake of 1990 and Bam earthquake of 2003 have proved that structures designed based on Iranian guidelines are seriously vulnerable against such near-fault earthquakes. Near-fault earthquakes are impulsive in nature, and structures are forced to withstand a huge amount of energy released in a short period of time. There are two main approaches to include the effects of such impulsive earthquakes in the design of infrastructures susceptible to such ground shakings. The first way is to modify the elastic acceleration response spectrum (EARS) in the linear based design method procedure. The next way is to conducting a nonlinear dynamic time history (NDTH) analysis for considering the effects of the impulsive nature of such quakes. In the fourth edition of Iranian seismic code (Standard No. 2800), several revisions have been made to take account of the effects of impulsive near-fault earthquakes. The modifications, proposed in the Standard No. 2800, are mainly based on the first way which is called spectral approach. In this study, several RC MRFs are designed linearly based on Standard No. 2800 and consequently, they are regenerated to become ready for nonlinear dynamic time history

(NDTH) analysis. The ultimate goal in this investigation is to find whether the modifications, which have been recently recommended based on spectral approach, are enough to consider the effects of such near-fault earthquakes using Performance Based Design (PBD) assessment methodology. In conclusion, the following results are found according to the analysis and evidences:

- In many cases, the seismic performance of low to mid-rise RC MRFs is not acceptable at all; all buildings are not almost able to maintain a life safety (LS) performance level. Therefore, in the case of nonlinear inter-story drift ratios (IDRs) and plastic hinge rotations (PHRs), structural demands surpass the LS limit states posed by guidelines such as ASCE41-06 or ASCE41-13 respectively. It is worthwhile to mention that structures designed based on design basis earthquakes (DBEs) are expected to meet LS performance level.
- In several cases, the structural demands, both IDRs and PHRs, exceed the collapse prevention (CP) performance level as well. While all buildings are designed based on DBE spectrum, it is unusual to see that they are not able to meet CP limit state. In this case, some structures are nearly at risk of getting collapsed.
- Analysis releases that taller RC MRFs, namely the twelve story buildings in this investigation, demonstrate much better seismic performance compared to the ones with shorter building's height.

In general, structures and infrastructures would not be allowed to be designed with Standard No. 2800 in regions susceptible to the severe impact of near-fault earthquakes. However, it would be wise to revise Standard No. 2800 to become suitable for the structures that are vital to be constructed near to an active fault. According to the abovementioned results, some recommendations seem useful to follow:

1- Regions and cities that are potentially at risk of being attacked by near-fault earthquakes must be detected and reported by highly qualified seismologists.
2- While structures are aimed to be designed by a linear based methods such as equivalent static procedure or modal analysis, site specific response spectrum, obtained from a seismic hazard analysis (SHA) of the neighbor active faults, should be used instead of using the standard response spectrum recommended by the Standard No. 2800.

3- Performing NDTH analysis must be obligatory for designing process of the structures if SHA could not be accomplished.

4- For conducting NDTH analysis, a bunch of near-fault ground motions compatibles with the soils condition of Iran must be selected by highly qualified seismologists. Consequently, they must become available for all structural engineers via a specific website recommended by the authorities of Standard No. 2800.

## Acknowledgments

The authors would like to sincerely acknowledge Dr. Saeid Pourzeynali, the Associate Professor of Guilan University, for his sincere supports, cooperation, motivations and advices. The authors also acknowledge valuable discussions on the research with M. Mashayekhi, M. Tootkaboni, and H. Moghaddam. At the end, the authors are particularly proud to dedicate this research to the people who lost their families and relatives in Manjil-Rudbar earthquake of 1990.

## References

Abrahamson, N.A. (2000). "Effects of rupture directivity on probabilistic seismic hazard analysis", Proceedings of the Sixth International Conference on Seismic Zonation, Earthquake Engineering Research Inst., Oakland, California.

Alavi, B., and Krawinkler, H. (2000) "Consideration of near-fault ground motion effects in seismic design." Proceedings, 12th World Conference on Earthquake Engineering, New Zealand, 1-8.

Anderson, J. C., and Bertero, V. V. (1987). "Uncertainties in establishing design earthquakes." ASCE Journal of Structural Engineering, 113(8), 1709-1724.

Baker, J. (2007). "Quantitative classification of near-fault ground motions using wavelet analysis." Bulletin of the Seismological Society of America, 97(5), 1486-1501.

Bray, J. D., and Rodriguez-Marek, A. (2004). "Characterization of forward-directivity ground motions in the near-fault region." Soil Dynamics and Earthquake Engineering, 24, 815-828.

Chopra, A. (1995). Dynamics of structures, Prentice Hall, Englewood Cliffs, NJ.

Chopra, A. K., and Chintanapakdee, C. (2001). "Comparing response of SDF systems to nearfault and far-fault earthquake motions in the context of spectral regions." Earthquake engineering and Structural Dynamics, 30, 1769-1789.

Hall, J. F., Heaton, T. H., Halling, M. W., and Wald, D. J. (1995). "Near-source ground motion and its effects on flexible buildings." Earthquake Spectra, 11(4), 569-605.

Makris, N. (1997). "Rigidity-plasticity-viscosity: can electro rheological dampers protect base isolated structures from near-source ground motions." Earthquake Engineering Structural and Dynamics, 26, 571–91.

Malhotra, P. K. (1999). "Response of Buildings to Near-Field Pulse-Like Ground Motions." Earthquake Engineering and Structural Dynamics, 28, 1309-1326.

Mavroeidis, G. P., and Papageorgiou, A. S. (2003). "A mathematical representation of near-fault ground motions." Bulletin of the Seismological Society of America, 93(3), 1099-1131.

McGuire, R. K. (2004). ''Seismic hazard and risk Analysis'', Earthquake Engineering Research Institute, Boulder, Colorado.

Mylonakis, G., and Reinhorn, A. (2001). "Yielding oscillator under triangular ground acceleration pulse." Journal of Earthquake Engineering, 5, 225-51. PEER. (1999). "Pacific Earthquake Engineering Research Center, strong motion database.

Priestley, M. J. N. (2003). ''Myths and fallacies in earthquake engineering'', revisited: The Malley-Milne lecture. Rose School, Collegio Allessandro Volta, Pavia, Italy.

Sasani, M., and Bertero, V. V. "Importance of severe pulse-type ground motions in performance based engineering: historical and critical review." Proceedings, 12th World Conference on Earthquake Engineering, New Zealand, 1-7.

Somerville, P. G. (2003). "Magnitude scaling of the near fault rupture directivity pulse." Physics of the Earth and Planetary Interiors, 137, 201-212.

Somerville, P. G., Smith, N. F., Graves, R., and Abrahamson, N. A. (1997). "Modification of Empirical Strong Ground Motion Attenuation Relations to Include the Amplitude and Duration Effects of Rupture Directivity." Seismological Research Letters, 68(1), 199-222.

Spudich, P., and Chiou, B. (2008). "Directivity in NGA earthquake ground motions: analysis using isochrone theory." Earthquake Spectra, 24(1), 279-98.

Zhang, Y., and Iwan, W. (2002). "Active interaction control of tall buildings subjected to near-field ground motions." Journal of Structural Engineering, 128, 69-79.

F. Naeim, S.E., H. Bhatia, P.E., Roy M. Lobo, P.E., (2000), ''Performance Based Seismic Engineering''. [book auth.] Farzad Naeim. The Seismic Design Handbook. Los Angeles, California : Springer.

Naeim, F. (1995). ''On seismic design implications of the 1994 Northridge earthquake record''. Earthquake Spectra 11:1, 91-109.

Hall, J.F., Heaton, T.H., Halling, M.W. and Wald, D.J. (1995). ''Near-source ground motion and its effects on flexible buildings''. Earthquake Spectra 11:4, 569-605.

Seismosoft Team (2013)" SeismoStruct User Manual (For version 7.2)." Pavia (PV). Italy: Seismosoft Ltd

Reza Sehhati, Adrian Rodriguez-Marek, Mohamed ElGawady, William F. Cofer (2011). ''Effects of near-fault ground motions and equivalent pulses on multi-story structures''. Engineering Structures 33 (2011) 767–779

N.N. Ambraseys, J. Douglas (2003). ''Near-field horizontal and vertical earthquake ground motions''. Soil Dynamics and Earthquake Engineering 23 (2003) 1–18

S. Yaghmaei-Sabegh, H.H. Tsangb (2011). ''An updated study on near-fault ground motions of the 1978 Tabas, Iran, earthquake (Mw=7.4)''. Scientia Iranica A (2011) 18 (4), 895–905

Varum, H. (2003). "Seismic Assessment, Strengthening and Repair of Existing Buildings", PhD Thesis, Department of Civil Engineering, University of Aveiro

Panagiotou. M, Restrepo J and Conte J (2011) "Shake-Table Test of a Full-Scale 7-Story Building Slice. Phase I: Rectangular"., s.l. : ASCE Journal of Structural Engineering, Vols. 137 (691-704). 0733-9445.

Mander JB, Priestley MJN and Park R (1988) "Theoretical stress-strain model for confined concrete," Journal of Structural Engineering, Vol. 114, No. 8, pp. 1804-1826

Menegotto M and Pinto PE (1973) "Method of analysis for cyclically loaded R.C. plane frames including changes in geometry and non-elastic behaviour of elements under combined normal force and bending,", International Association for Bridge and Structural Engineering, Zurich, Switzerland, pp. 15-22

American Concrete Institute (ACI). (2011). "Building code requirements for structural concrete." ACI 318-11, Farmington Hills, MI.

Hilber H.M., Hughes T.J.R., Taylor R.L. (1977) "Improved numerical dissipation for time integration algorithms in structural dynamics," Earthquake Engineering and Structural Dynamics, Vol. 5, No. 3, pp. 283-292.

Priestley M.J.N., Grant D.N. (2005) "Viscous damping in seismic design and analysis," Journal of Earthquake Engineering, Vol. 9, Special Issue 1, pp. 229-255.

Moghaddam.H, Harati.M, (2015), "Seismic performance comparison of mid-rise moment resisting frame and shear wall system", Proceedings of the seventh International Conference on Seismology & Earthquake Engineering (SEE7), Tehran, IRAN.

Federal Emergency Management Agency (FEMA) (2009). Recommended Methodology for Quantification of Building System Performance and Response Parameters, FEMA P695, Prepared for the Federal Emergency.